# First-principles study of structural and electronic properties of multiferroic oxide Mn$_3$TeO$_6$ under high pressure

Xiao-Long Pan(潘小龙)[1,2], Hao Wang(王豪)[2], Lei Liu(柳雷)[2], Xiang-Rong Chen(陈向荣)[1]*, Hua-Yun Geng(耿华运)[2,3]*

[1] *College of Physics, Sichuan University, Chengdu 610065, P. R. China;*

[2] *National Key Laboratory of Shock Wave and Detonation Physics, Institute of Fluid Physics, China Academy of Engineering Physics, Mianyang, Sichuan 621900, P. R. China;*

[3] *HEDPS, Center for Applied Physics and Technology, and College of Engineering, Peking University, Beijing 100871, P. R. China;*

**Abstract:** Mn$_3$TeO$_6$ (MTO) has been experimentally found to adopt a *P2$_1$/n* structure under high pressure, which exhibits a significantly smaller band gap compared to the atmospheric *R$\bar{3}$* phase. In this study, we systematically investigate the magnetism, structural phase transition and electronic properties of MTO under high pressure through first-principles calculations. Both *R$\bar{3}$* and *P2$_1$/n* phases of MTO are antiferromagnetic at zero temperature. The *R$\bar{3}$* phase transforms to the *P2$_1$/n* phase at 7.58 GPa, accompanied by a considerable volume collapse of about 6.47%. Employing the accurate method that combines DFT+$U$ and G$_0$W$_0$, the calculated band gap of *R$\bar{3}$* phase at zero pressure is very close to the experimental values, while that of the *P2$_1$/n* phase is significantly overestimated. The main reason for this difference is that the experimental study incorrectly used the Kubelka-Munk plot for the indirect band gap to obtain the band gap of the *P2$_1$/n* phase instead of the Kubelka-Munk plot for the direct band gap. Furthermore, our study reveals that the transition from the *R$\bar{3}$* phase to the *P2$_1$/n* phase is accompanied by a slight reduction in the band gap.

* *Corresponding authors. E-mail: xrchen@scu.edu.cn; s102genghy@caep.cn*

## I. Introduction

In recent years, multiferroic oxides have been extensively investigated for their potential photovoltaic applications[1-3]. As potential alternatives for the metal-halide perovskite solar cells, multiferroic oxides overcome the disadvantages (e.g., degradation, hysteresis, and lead pollutions) of traditional halide perovskites[4-7]. However, multiferroic oxides typically have wide band gaps (2.7–4.0 eV)[8] that exceed the ideal value of 1.4 eV (corresponding to the maximum theoretical power conversion efficiency (PCE) for a photovoltaic material)[9], which hinders further improvement of their PCE. Therefore, reducing the band gaps of multiferroic oxides is an effective strategy to obtain materials with a high PCE. Double perovskite multiferroic oxide $Bi_2CrFeO_6$ has been shown to have a PCE of up to 8.1% by engineering the cationic ordering[2]. In addition, pressure, as an effective way to regulate the band gap, has been applied to the multiferroic oxide $Mn_3TeO_6$ (MTO)[3].

Recently, Liu et al. discovered that the band gap of multiferroic oxide MTO ($R\bar{3}$) was reduced by 39% (from 3.15 eV to 1.86 eV) under compression with a structural phase transition, corresponding to a transition pressure between 16 and 18 GPa[3]. Such a reduction of the band gap is beneficial for applications in light-absorbing materials, which usually possess a small bandgap of about 1.4 eV. Subsequently, Arévalo-López et al. synthesized a novel double perovskite structure of MTO by subjecting $R\bar{3}$ crystals to high-pressure and high-temperature conditions (8 GPa and 1173 K)[10]. Their research also revealed a decrease of the band gap from 2.60 eV of the $R\bar{3}$ phase to 1.80 eV of the $P2_1/n$ phase, with a reduction of 0.80 eV, and they suggest that this reduction explains the significant decrease in MTO's band gap under high pressure, as observed by Liu et al. At almost the same time, Su et al. reported that phase transition from the $R\bar{3}$ phase to the $P2_1/n$ phase in MTO occurs at the pressure of about 5 GPa based on first-principles calculations and experimental results by treating the MTO ($R\bar{3}$) crystals at 5 GPa and 1173 K[11].

Although there is sufficient evidence in the existing studies to show that the $P2_1/n$ phase can be a high phase of the MTO, the phase transition pressure from the $R\bar{3}$ structure to the *P2₁/n* structure reported by Su et al. is significantly lower than the phase transition pressure range of 16-18 GPa reported by Liu et al. In order to elucidate the underlying mechanism behind the significant reduction of the band gap observed in MTO, Liu et al. further conducted in-situ X-ray diffraction analysis on the structural evolution of MTO during compression and decompression processes, discovering a sequence of irreversible phase transitions: $R\bar{3}$→*C2/c*→*P2₁/n*[12]. Under compression, $R\bar{3}$ transforms into a *C2/c* structure at 19 GPa, and under decompression, the *C2/c* structure transforms into a *P2₁/n* structure at about 9.6 GPa. Currently, these studies have indicated that the *P2₁/n* structure is a high-pressure phase of MTO. However, considering the differences in transition pressure from the $R\bar{3}$ phase to the *P2₁/n* phase in previous studies, and to ascertain whether the phase with a significantly reduced band gap in MTO is indeed the *P2₁/n* structure, further studies are needed to confirm this.

In this study, to theoretically validate the high-pressure phase transition of MTO and its associated significant reduction of the band gap, we systematically investigated the magnetic properties, phase transition, and electronic properties of MTO under high pressure through first-principles calculations. It is worth noting that compared with previous studies, we further studied the high-pressure phase transition and band gap of MTO in detail. A more accurate method combining DFT+*U* and G$_0$W$_0$ is used to calculate the band gap of MTO. In next section, the computational method is formulated, the results and discussion are presented in section III, and section IV summarizes the conclusions of the paper.

## II. Computational Method

All first-principles calculations were performed using density functional theory (DFT)[13, 14] as implemented in the Vienna Ab initio Simulation Package (VASP)[15, 16]. The description of the exchange-correlation functional adopted the Perdew-Burke-Ernzerhof (PBE)[17] generalized gradient approximation (GGA). The atomic coordinates and lattice vectors were fully optimized until the Hellmann–Feynman forces acting on each atom less than $10^{-2}$ eV/Å and the convergence of the total energy was better than $10^{-6}$ eV between two successive ionic steps. The cut-off energy for the plane-wave basis was set to 550 eV, and Monkhorst-Pack[18] k-meshes with a grid spacing of $2\pi \times 0.02$ Å$^{-1}$ were used to sample the first irreducible Brillouin zone for all calculations, ensuring that the energy for each structure converges to 1 meV per atom. Spin polarization is taken into account in our calculations. The cell shape is completely relaxed in the calculations of structural relaxation to achieve the hydrostatic conditions. The Hubbard correction was included in the GGA+$U$ approximation to account for strong correlation interaction between the 3$d$ electrons of Mn atoms. Specifically, we used the simplified GGA+$U$ method proposed by Dudarev et al.[19] with only a single effective $U$ parameter. The value of $U$ is set to 4 eV as usually adopted for Mn oxides, and we also consider some other $U$ values for comparison in the calculations of phase transition and band gap.

In order to ensure the dynamic stability of the $R\bar{3}$ and $P2_1/n$ phases, the phonon dispersion curves were calculated using the finite displacement method implemented in ALAMODE code[20]. A $2 \times 2 \times 2$ supercell for the $R\bar{3}$ phase (160 atoms) and $1 \times 1 \times 1$ supercell for the $P2_1/n$ phase (40 atoms) were employed in the phonon spectra calculations. Given that the phonon spectrum of the $P2_1/n$ phase at zero temperature exhibits slight imaginary frequency, we calculated its phonon spectrum at high temperature by using the self-consistent phonon theory. The displacements and forces required to fit the anharmonic force constants were extracted from the snapshots obtained during *ab initio* molecular dynamics (AIMD) simulation. We conducted AIMD simulation of the $P2_1/n$ phase at 300 K using canonical ensemble, comprising a total of 4000 steps with a time step of 1.5 fs. Under equilibrium conditions, we extracted 100 sets of displacements and forces from the final 2000 steps, with an interval of 10 steps.

As is well-known, the GGA exchange-correlation function tends to underestimate the band gaps of semiconductors containing 3$d$ electrons. Therefore, it is necessary to employ a more accurate quasiparticle approximation GW approach to calculate the band gap of MTO. In our calculations, on the basis of GGA+$U$ wave functions, $G_0W_0$ calculation is adopted to obtain the exact quasiparticle band structures. Wannier interpolation is used to process the calculation results of $G_0W_0$ to obtain quasiparticle band structures in Wannier90 software[21].

## III. Results and discussion

The four alternative phases of MTO considered in our calculations are the $R\bar{3}$ structure (existing at atmospheric conditions), the $P2_1/n$ structure (confirmed to be stable at low pressure), the $Ni_3TeO_6$-type $R3$ structure (common in transition metal double perovskites), and the fourth structure, the β-$Li_3VF_6$-type $C2/c$ structure (which was found to be a high-pressure structure of MTO[12]). As shown in Fig. 1, the $R\bar{3}$ and $R3$ phases have a triangular structure, while the $P2_1/n$ and $C2/c$ phases have a monoclinic structure. In these structures, O atoms are distributed around Mn and Te atoms to form octahedra, $TeO_6$ octahedra are very regular, whereas $MnO_6$ octahedra are considerably distorted.

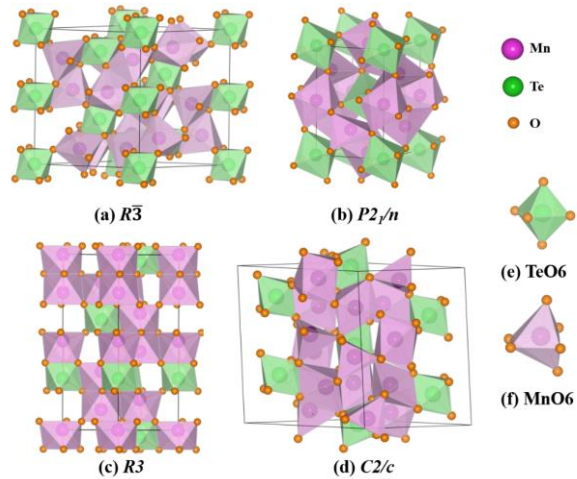

**Fig. 1** (color online) Four crystal structures of $Mn_3TeO_6$ considered in our calculations. (a) $R\bar{3}$, (b) $P2_1/n$, (c) $R3$, and (d) $C2/c$, as well as (e) $TeO_6$ octahedra and (f) distorted $MnO_6$ octahedra.

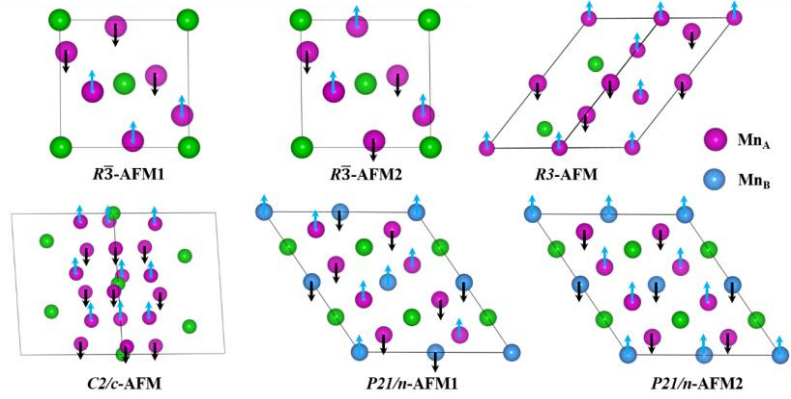

**Fig. 2** (color online) Candidate antiferromagnetic ordering configurations for $R\bar{3}$, $P2_1/n$, $R3$, and $C2/c$ phases of MTO considered in our calculations. To facilitate comparison with the experimental data, the Mn atoms in the $P2_1/n$ phase can be classified into $Mn_A$ and $Mn_B$ atoms based on their distinct symmetries, with $Mn_A$ occupying the $4e$ sites and $Mn_B$ occupying the $2b$ sites of the lattice.

For these structures, we first compare the energy differences between different magnetic configurations to find the most stable magnetic configuration, including antiferromagnetic and ferromagnetic configurations. The candidate antiferromagnetic ordering configurations of the $R\bar{3}$, $P2_1/n$, $R3$, and $C2/c$ structures are displayed in Fig. 2. It is noteworthy that, for the $R\bar{3}$ structure, we used a rhombohedral cell with 20 atoms for magnetic calculations. For the $P2_1/n$ structure, we adopted the magnetic cell with 40 atoms proposed by Arévalo-López et al., obtained using the transformation matrix: [(1 0 -1), (0 -1 0), (-2 0 0)] with an origin shift of (0, 0, 0.5). For the $R3$ structure, we use the $1 \times 1 \times 2$ supercell of its primary cell as the magnetic structure. For the $R\bar{3}$ and $P2_1/n$ phases, we consider two antiferromagnetic configurations, while for the $R3$ and $C2/c$ phases, we only consider the simplest one. Using these magnetic configurations, we calculated the energy-volume curves for the $R\bar{3}$, $P2_1/n$, $R3$, and $C2/c$ phases, respectively, as shown in Fig. 3. We can see that the antiferromagnetic configuration is more stable than the ferromagnetic state for all phases. For $R\bar{3}$ phase, AFM1 configuration is obviously the most stable, while for $P2_1/n$, the energy of AFM1 and AFM2 orders is almost the same, with the AFM2 order slightly lower than the AFM1 order. The AFM2 configuration of the $P2_1/n$ phase obtained by our calculation is consistent with the magnetic configuration proposed by Arévalo-López et al.[10]. Using the most stable magnetic configuration for these structures, we calculated their lattice constants at zero pressure, as illustrated in Table 1. The lattice parameters of these structures calculated in our study are in close agreement with experimental values, differing by less than 2%.

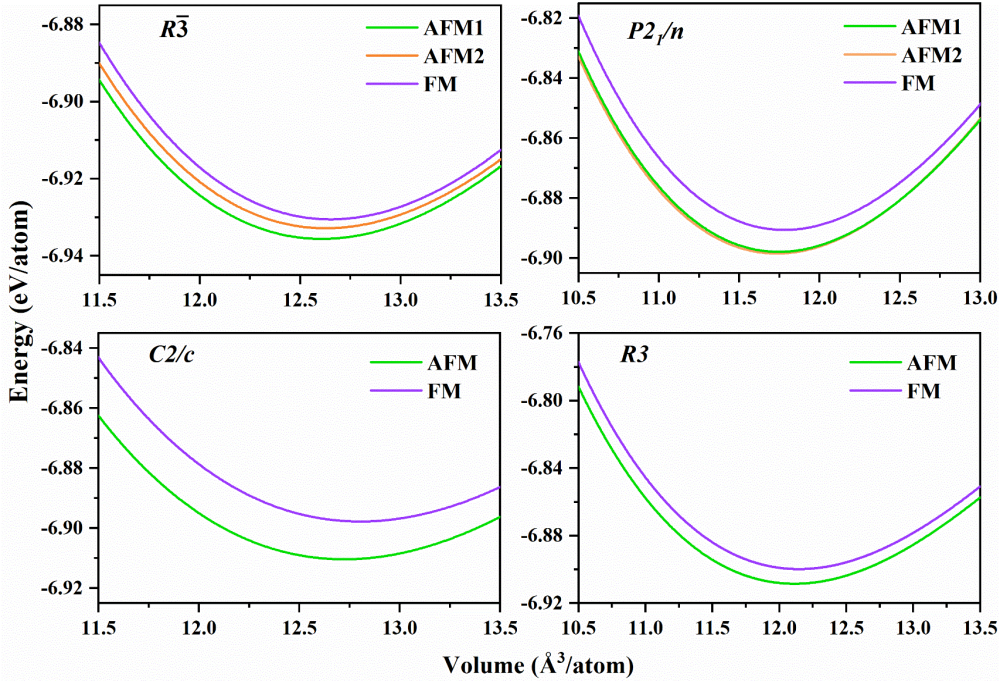

**Fig. 3** (color online) Calculated energy-volume curves of the $R\bar{3}$, $P2_1/n$, $R3$, and $C2/c$ phases with the different magnetic orders, respectively.

**Table 1** The calculated lattice parameters of the $R\bar{3}$, $P2_1/n$, $R3$, and $C2/c$ phases by DFT+$U$ method with $U$=4 eV, along with previous experimental data for comparison.

| Structure | $a$ (Å) | $b$ (Å) | $c$ (Å) | $\alpha$(°) | $\beta$(°) | $\gamma$(°) |
|---|---|---|---|---|---|---|
| $R\bar{3}$ (this work) | 8.94 | 8.94 | 10.77 | 90 | 90 | 120 |
| $R\bar{3}$ (Su et al.) | 8.90 | 8.90 | 10.71 | 90 | 90 | 120 |
| $R\bar{3}$ (Ivanov et al.) | 8.87 | 8.87 | 10.67 | 90 | 90 | 120 |
| $P2_1/n$ (this work) | 5.35 | 5.49 | 7.87 | 90 | 89.77 | 90 |
| $P2_1/n$ (Su et al.) | 5.29 | 5.45 | 7.80 | 90 | 89.63 | 90 |
| $P2_1/n$ (Arévalo-López et al.) | 5.29 | 5.45 | 7.81 | 90 | 89.63 | 90 |
| $R3$ (this work) | 5.41 | 5.41 | 14.21 | 90 | 90 | 120 |
| $C2/c$ (this work) | 8.96 | 8.96 | 10.74 | 92.99 | 92.99 | 61 |

After obtaining the most stable magnetic configuration of each structure, we further investigate the stability of these structures under high pressure to explore the structural phase transition of MTO. The calculated enthalpy difference of the *P2₁/n*, *R3*, and *C2/c* structures relative to the $R\bar{3}$ is shown in Fig. 4 (a). The $R\bar{3}$ phase is the most stable phase in the pressure range of 0~7.58 GPa. A pressure-driven phase transition from $R\bar{3}$ phase to *P2₁/n* phase occurs at the pressure of 7.58 GPa for MTO, consistent with the results of Su et al.[11]. Different *U* values selected in the GGA+*U* calculations have a negligible impact on the phase transition pressure of MTO, which increases slightly with larger *U* values (Fig. 4(b)). Specifically, the phase transition pressure calculated with a *U* value of 2 eV is only 0.13 GPa lower than the primarily used *U* value of 4 eV in our calculations. Above 7.58 GPa, the *P2₁/n* phase remains stable up to 30 GPa (the maximum pressure we investigated), whereas *C2/c* structure is energy unfavorable throughout the pressure area (~0.03 eV/atom higher enthalpy than the $R\bar{3}$ structure at 30 GPa). Our theoretical result is inconsistent with previous experimental study that identified a structural phase transition from $R\bar{3}$ to *C2/c* at 19 GPa. This discrepancy may arise from the considerably lower energy barrier for the transition from $R\bar{3}$ to *C2/c* compared to the transition from $R\bar{3}$ to *P2₁/n*, emphasizing the crucial role of kinetic effects in these transitions[12]. In addition, the $R\bar{3}$ to *P2₁/n* transition of MTO is a typical first-order phase transition with a significant volume reduction of about 6.47%, as depicted by the equation of state (EOS) in Fig. 5. By fitting calculated EOS of the $R\bar{3}$ phase to the Vinet EOS[22], we obtained that the volume modulus ($B_0$) of MTO at zero pressure is 126.16 GPa, and its pressure derivative (B') is 4.13.

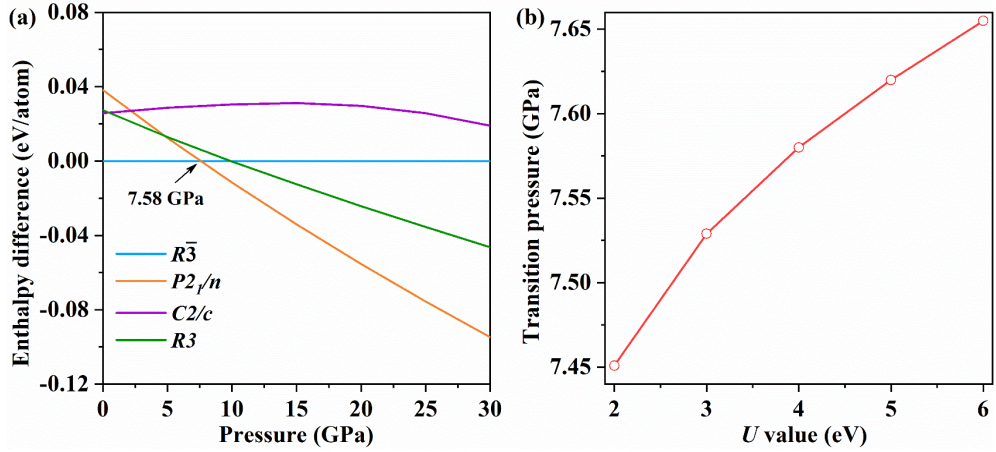

**Fig. 4** (color online) (a) The calculated enthalpy difference of $P2_1/n$, *R3*, and *C2/c* structures with respect to $R\bar{3}$ structure. (b) The change of phase transition pressure with different *U* values in the DFT+*U* calculations.

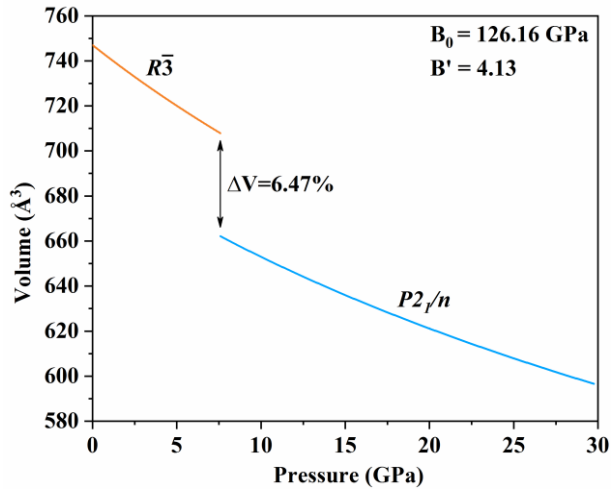

**Fig. 5** (color online) Calculated equation of state (EOS) for MTO. For comparison, the volume values of both phases are standardized to the volume corresponding to *Z*=6.

Furthermore, the calculated phonon spectra further confirm the dynamic stability of the $R\bar{3}$ and $P2_1/n$ phases at zero pressure. The phonon spectrum of the $R\bar{3}$ phase in Fig. 6(a) shows that it is dynamically stable. Given the slight imaginary frequency observed in the phonon spectrum of the $P2_1/n$ phase at zero temperature, we proceeded to calculate its phonon spectrum at 300 K using the self-consistent phonon theory, which demonstrates its dynamic stability at room temperature, as shown in Fig. 6(b).

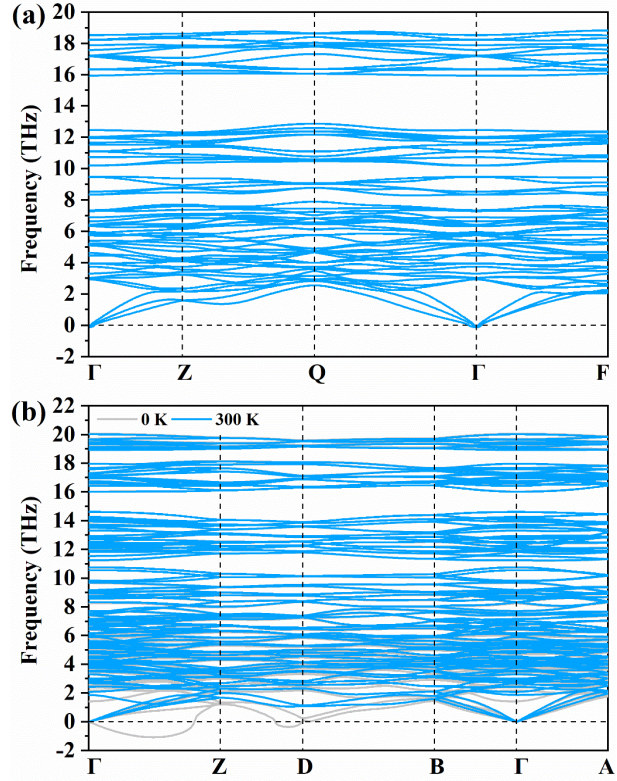

Fig. 6 (color online) The calculated phonon spectra of the (a) $R\bar{3}$ and (b) $P2_1/n$ phases. For $P2_1/n$ phase, the phonon spectrum at 300 K was calculated by self-consistent phonon theory.

Next, we examine the most striking electronic properties of MTO. Results of GGA+$U$ calculations reveal that both $R\bar{3}$ and $P2_1/n$ are semiconductors with direct band gap, and their band gap is 1.275 eV and 1.056 eV at zero pressure, respectively, as shown in Fig. 7. Note that since the spin-up and spin-down band structures for both the $R\bar{3}$ and $P2_1/n$ phases are the same, we have only presented the spin-up band structures. Furthermore, pressure dependence of band gap for MTO is illustrated in Fig. 8(a). In the stable pressure region of $R\bar{3}$ and $P2_1/n$ phases, their band gaps increase with higher pressure, and there is a decrease in the band gap by 0.197 eV (14.8%) at the phase transition pressure of 7.58 GPa. Additionally, we also investigate the relationship between the band gap of $R\bar{3}$ and $P2_1/n$ phases and the $U$ value parameter of DFT+$U$ calculations, with the results presented in Fig. 8(b). The band gaps of $R\bar{3}$ and $P2_1/n$ phases increase with the increased $U$ value. When the calculations using pure GGA method or a small $U$ value will incorrectly predict $P2_1/n$ phase to be metallic; on the contrary, a large $U$ value ($\geq$ 7 eV) shows that the band gaps of $R\bar{3}$ and $P2_1/n$ phases are almost equal at the phase transition pressure, which is inconsistent with the previous experimental studies. The calculated results with a $U$ value of 4 eV show that both the band gaps and the band gap difference of the two phases are moderate, further confirming the rationality of our selected $U$ value.

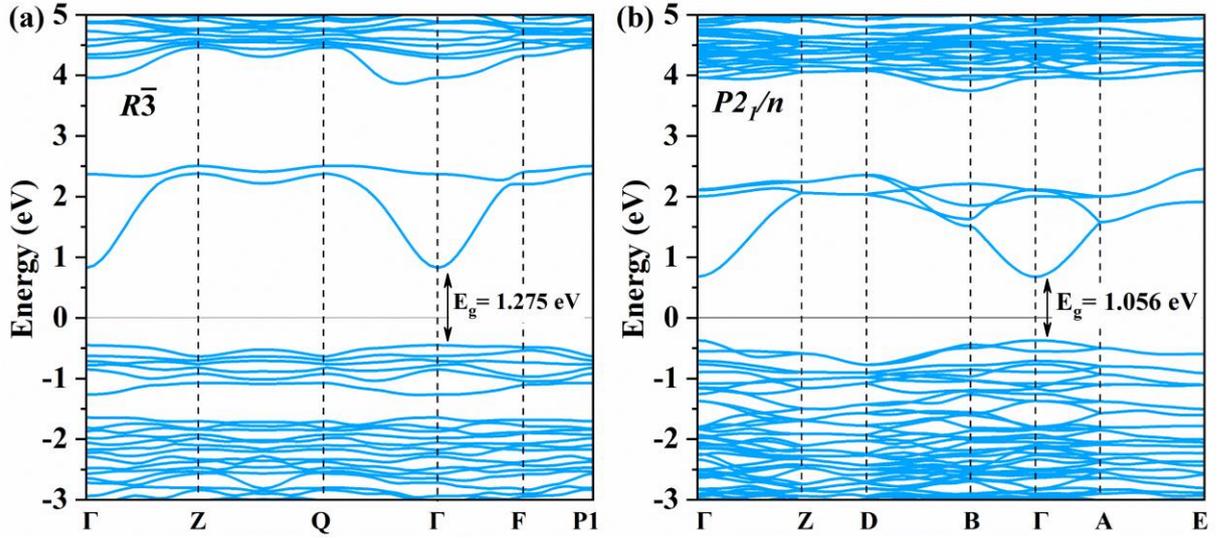

**Fig. 7** Electronic band structures of the $R\bar{3}$ and $P2_1/n$ phases at zero pressure calculated by DFT+$U$ approach.

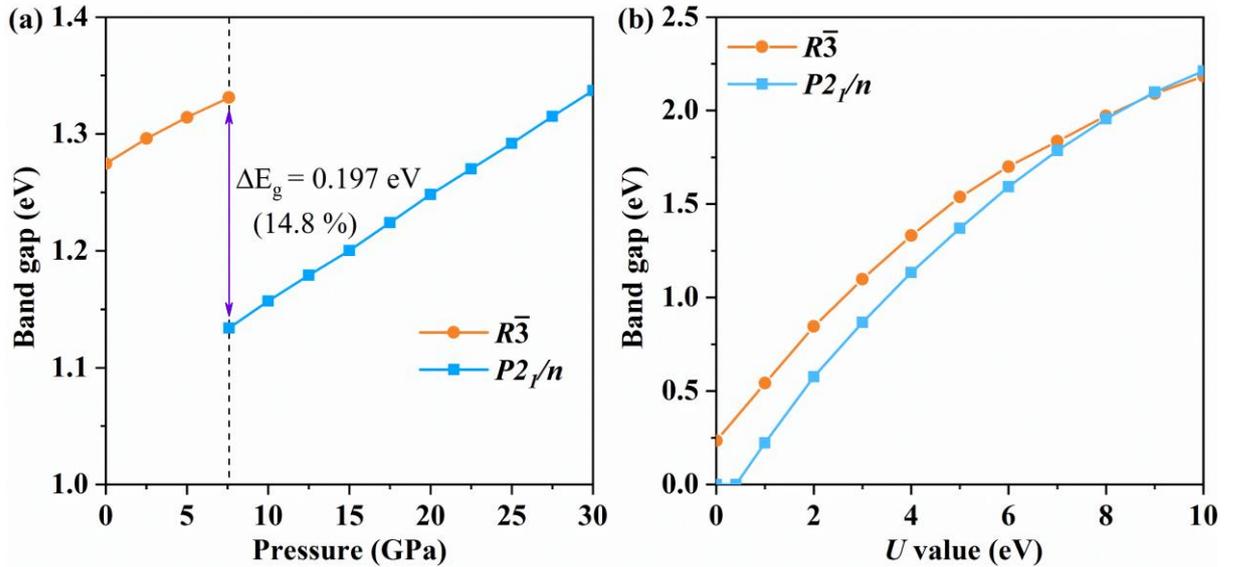

**Fig. 8** (color online) (a) Pressure dependence of band gap for MTO calculated by DFT+$U$ approach with $U$=4 eV. (b) Band gaps of the $R\bar{3}$ and $P2_1/n$ phases at the phase transition pressure as a function of the $U$ value.

Given that the electronic band structures calculated by GGA+$U$ method underestimate the band gap compared to the experimental value, it is essential to employ a more accurate method to reproduce the experimental band gap of MTO. It's well known that many-body perturbation theory within the GW approximation has achieved great success in predicting the quasiparticle band structure of solid materials. Here, we combine DFT+$U$ and $G_0W_0$ methods to calculate the quasiparticle band structures of MTO, and the corresponding results are shown in Fig. 9. Both $R\bar{3}$ and $P2_1/n$ phase are direct band gap semiconductor with large band gaps of 2.836 eV and 2.768 eV, respectively, and the lowest point of conduction band energy and the highest point of valence band energy are both at the gamma point. Compared with the simple GGA+$U$ method, the band gaps calculated by $G_0W_0$ method have an obvious increase, with an increment of 1.561 eV for $R\bar{3}$ and 1.712 eV for $P2_1/n$. In addition, electronic projected density of states (PDOS) of MTO shows that near the Fermi level the conduction band is mainly contributed by the $2p$ orbitals of O atoms and the $5s$ orbitals of Te atoms for both $R\bar{3}$ and $P2_1/n$ phases, while the valence band is mainly contributed by the $3d$ orbitals of Mn atoms and the $2p$ orbitals of O atoms.

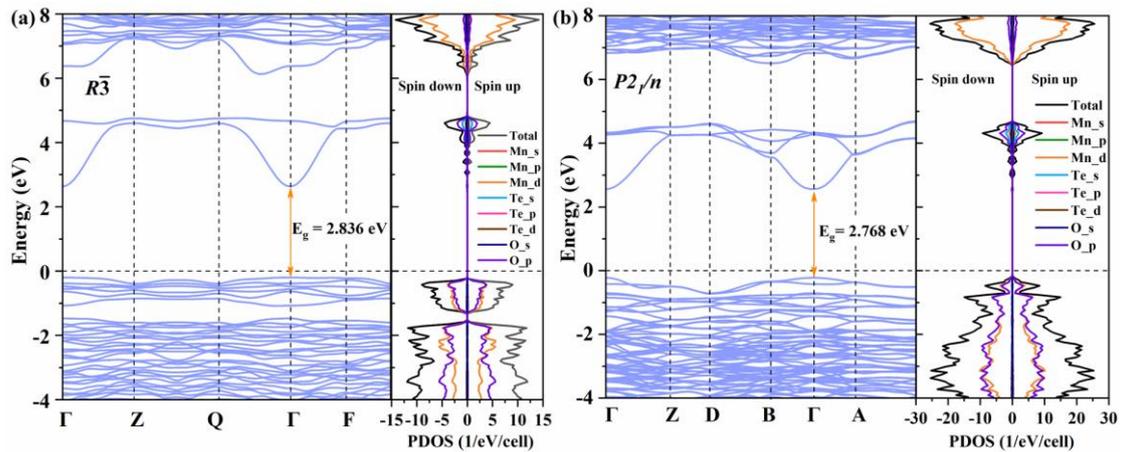

**Fig. 9** (color online) The quasiparticle band structures and electronic projected density of states (PDOS) of the $R\bar{3}$ and $P2_1/n$ phases at 0 GPa calculated by combining GGA+$U$ and $G_0W_0$ approach.

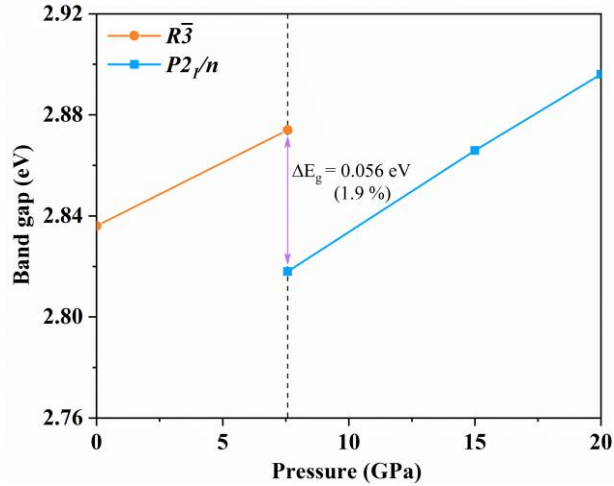

**Fig. 10** (color online) Pressure dependence of band gap for MTO calculated by combining GGA+$U$ and $G_0W_0$ approach.

Furthermore, the calculated band gaps of MTO under high pressure in Fig. 10 shows that the band gap reduction from $R\bar{3}$ and $P2_1/n$ phases is only 0.056 eV at the phase transition pressure of 7.58 GPa, which is significantly lower than 0.8 eV observed by Arévalo-López et al.[10]. Although the calculated band gap of 2.836 eV for the $R\bar{3}$ phase at zero pressure is close to the previously measured values of 2.6 eV by Arévalo-López et al. and approximately 3.05 eV by Liu et al., it must be acknowledged that our calculations overestimate the band gap of the $P2_1/n$ phase by about 1 eV compared to the experimental value of 1.8 eV reported by Arévalo-López et al.[10]. The reason why our calculated band gap of the $P2_1/n$ phase is significantly higher than the experimental value reported by Arévalo-López et al. is that they treat the $P2_1/n$ phase as an indirect band gap semiconductor when using Kubelka-Munk plot to fit the band gap, as shown in Fig. 11(a). However, our calculations reveal that the $P2_1/n$ phase is a direct bandgap semiconductor, and Liu et al.'s experimental treatment[3] also categorizes MTO as a direct bandgap semiconductor. According to the Kubelka-Munk formula:

$$(F(R)h\nu)^{1/n} = B\left(h\nu - E_g\right),$$

where $R$ is the reflectance, $h$ is the Planck constant, $\nu$ is the frequency of the light, and the value of the constant $n$ is 2 for indirect bandgap semiconductors and 1/2 for direct bandgap semiconductors, we reprocessed the experimental UV-VIS reflectance spectra data of Arévalo-López et al. using the formula for the direct band gap ($n=1/2$), and the new Kubelka-Munk plot is shown in Fig. 11 (b). The Kubelka-Munk plot for a direct bandgap has an obvious linear segment in the energy range of 2.75~3.25 eV, and the linear fitting result shows that the band gap of the $P2_1/n$ phase is about 2.7 eV, which is very close to our calculated value. In addition, the linear segment of the Kubelka-Munk plot for a direct bandgap is more obvious than that for an indirect bandgap, which further confirms that the Kubelka-Munk plot for a direct bandgap should be adopted to obtain the band gap of the $P2_1/n$ phase. By the way, the calculated magnetic moments for $Mn_A$ and $Mn_B$ atoms in the $P2_1/n$ phase are 4.49 $\mu_B$ and 4.46 $\mu_B$, respectively. While our calculated average magnetic moment per Mn atom of 4.48 $\mu_B$ matches the experimental value of 4.47 $\mu_B$, a difference arises in the comparison with experimental values of 4.8 $\mu_B$ for $Mn_A$ atoms and 3.8 $\mu_B$ for $Mn_B$ atoms[10]. We further consider the effects of spin-orbit coupling (SOC) and non-collinear magnetism. The calculated results indicate that the magnitude of the magnetic moments of Mn atoms in the $P2_1/n$ phase hardly changes. Only the direction of the magnetic moment deviates slightly from the z-axis direction, with a maximum deviation angle of less than 1.5°.

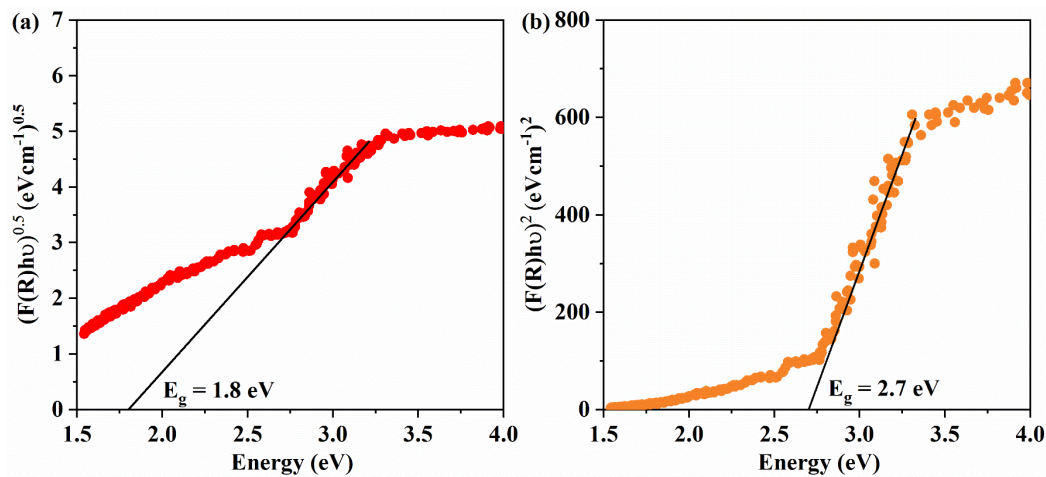

**Fig. 11** (color online) Kubelka–Munk plots of the $P2_1/n$ phase for (a) an indirect band gap ($n=2$) and (b) a direct band gap ($n=1/2$). The corresponding UV-VIS reflectance spectra data of the $P2_1/n$ phase are extracted from Figure 4 in the article of Arévalo-López et al[10].

## IV. Conclusion

Through our first-principles GGA+$U$ calculations, we determined that the antiferromagnetic configuration of the $R\bar{3}$ phase of MTO is more stable than the ferromagnetic configuration. The $P2_1/n$ phase of MTO is further confirmed to be stable under high pressure, and exhibits antiferromagnetism, consistent with previous experimental studies. At 7.58 GPa, the $R\bar{3}$ phase undergoes a first-order phase transition, transforming into the $P2_1/n$ phase, accompanied by a considerable reduction in atomic volume of about 6.47%.

The band gap of the $R\bar{3}$ phase at zero pressure, calculated using the DFT+$U$ method, is determined to be 1.275 eV. This value is notably lower than the experimental values of 2.60 eV and 3.05 eV. Employing a more accurate method that combines DFT+$U$ and $G_0W_0$, we obtained a band gap of 2.836 eV for the $R\bar{3}$ phase, which is almost consistent with the experimental values. However, our study shows that the calculated band gap (2.768 eV) of the $P2_1/n$ phase significantly overestimates the experimental value of 1.80 eV by approximately 1 eV, thereby not to theoretically predict the substantial reduction in the band gap from the $R\bar{3}$ phase to the $P2_1/n$ phase observed in previous experimental study. The reason why our theoretically calculated band gap of the $P2_1/n$ phase is significantly higher than the experimental value is that the experimental band gap was obtained using the Kubelka-Munk plot for the indirect band gap. However, the $P2_1/n$ phase is actually a direct bandgap semiconductor. Its band gap should be determined using the Kubelka-Munk plot for the direct band gap. The band gap value of the $P2_1/n$ phase obtained by the Kubelka-Munk plot for the direct band gap is close to our calculated value. Our results also suggest that the phase transition from the $R\bar{3}$ phase to the $P2_1/n$ phase is accompanied by only a slight reduction in the band gap.

## Acknowledgments

This work was supported by National Key R&D Program of China under Grant No. 2021YFB3802300, the NSAF under Grant Nos. U1730248 and U1830101, the

National Natural Science Foundation of China under Grant Nos.12202418, 11872056, 11904282, 12074274, and 12174356.

## Author Contribution

Xiao L. Pan: Calculation, Analysis, Writing, Original draft preparation. Hao Wang: Analysis, Reviewing and Editing. Lei Liu: Analysis, Writing, Reviewing and Editing. Xiang R. Chen: Writing, Reviewing and Editing, Fund. Hua Y. Geng: Project design, Writing, Reviewing and Editing, Fund.